%% file: main.tex
\documentclass[
]{ceurart}

\usepackage{enumitem}
\usepackage[shortcuts]{extdash}
\usepackage[all]{foreign}
\usepackage{cleveref}
\usepackage{subcaption}
\usepackage{tablefootnote}

\usepackage{newfloat}
\usepackage[frozencache,cachedir=.]{minted}
\setminted{style=bw}
\AtBeginEnvironment{minted}{\vspace{-4pt}}
\AtEndEnvironment{minted}{\vspace{-4pt}}

\usepackage{glossaries}
\glsdisablehyper
\input{acronyms}
\usepackage{array}
\usepackage{amssymb}
\usepackage{pifont}
\usepackage[ruled,vlined]{algorithm2e}
\SetAlFnt{\footnotesize}
\SetAlCapFnt{\small}
\SetAlCapNameFnt{\small}

\newcommand{\BibTeX}{B\kern-.05em{\sc i\kern-.025em b}\kern-.08em\TeX}
\newcommand{\footurl}[1]{\footnote{\url{#1} (on August 6, 2025).}}
\newcommand{\tablefooturl}[1]{\tablefootnote{\url{#1} (on August 6, 2025).}}

\begin{document}

%%
%% Rights management information.
%% CC-BY is default license.
\copyrightyear{2025}
\copyrightclause{Copyright for this paper by its authors.
  Use permitted under Creative Commons License Attribution 4.0
  International (CC BY 4.0).}

%%
%% This command is for the conference information
\conference{XAI-KRKG@ECAI25: First International ECAI Workshop on eXplainable AI, Knowledge Representation and Knowledge Graphs, October 25–30, 2025, Bologna, Italy}

%%
%% The "title" command
\title{Enabling Transparent Cyber Threat Intelligence Combining Large Language Models and Domain Ontologies}

%%% Use this combinations of commands to specify all authors of your 
%%% paper. Use \fnms{} and \snm{} to indicate everyone's first names 
%%% and surname. This will help the publisher with indexing the 
%%% proceedings. Please use a reasonable approximation in case your 
%%% name does not neatly split into "first names" and "surname".
%%% Specifying your ORCID digital identifier is optional. 
%%% Use the \thanks{} command to indicate one or more corresponding 
%%% authors and their email address(es). If so desired, you can specify
%%% author contributions using the \footnote{} command.

\author[1]{Luca, Cotti}[orcid=0009-0004-6351-556X,email= luca.cotti@unibs.it]
\cormark[1]
\author[1]{Anisa, Rula}[orcid=0000-0002-8046-7502]
\author[1]{Devis, Bianchini}[orcid=0000-0002-7709-3706]
\author[1,2,3]{Federico, Cerutti}[orcid=0000-0003-0755-0358]

\address[1]{Department of Information Engineering, University of Brescia, Italy}
\address[2]{School of Computer Science and Informatics, Cardiff University, United Kingdom}
\address[3]{Department of Electronics and Computer Science, University of Southampton, United Kingdom}

%% Footnotes
\cortext[1]{Corresponding author.}

%%
%% The abstract is a short summary of the work to be presented in the
%% article.
\input{parts/abstract}

%%
%% Keywords. The author(s) should pick words that accurately describe
%% the work being presented. Separate the keywords with commas.

%%
%% This command processes the author and affiliation and title
%% information and builds the first part of the formatted document.
\maketitle

\section{Introduction}

\input{parts/introduction}

\section{Background}
\label{sec:background}

\input{parts/background}

\section{Methodology}
\label{sec:methodology}

\input{parts/methodology}

\section{Experimental Results}
\label{sec:experiment}

\input{parts/experiments}

\section{Conclusions}
\label{sec:conclusions}
\input{parts/conclusions}

%%
%% The acknowledgments section is defined using the "acknowledgments" environment
%% (and NOT an unnumbered section). This ensures the proper
%% identification of the section in the article metadata, and the
%% consistent spelling of the heading.
\begin{acknowledgments}
This work was partially supported by project SERICS (PE00000014) under the MUR National Recovery and Resilience Plan funded by the European Union – NextGenerationEU, specifically by the project NEACD: Neurosymbolic Enhanced Active Cyber Defence (CUP J33C22002810001). This project was also partially funded by the Italian Ministry of University as part of the PRIN: PROGETTI DI RICERCA DI RILEVANTE INTERESSE NAZIONALE – Bando 2022, Prot. 2022EP2L7H
\end{acknowledgments}

%% The declaration on generative AI comes in effect
%% in Janary 2025. See also
%% https://ceur-ws.org/GenAI/Policy.html
\section*{Declaration on Generative AI}
During the preparation of this work, the authors used ChatGPT-4o for the following purposed: Grammar and spelling check, Paraphrase and reword. After using this tool, the authors reviewed and edited the content as needed and take full responsibility for the publication’s content. 

\bibliography{biblio.bib}

\end{document}

%% file: acronyms.tex
\newacronym{ai}{AI}{Artificial Intelligence}
\newacronym{cti}{CTI}{Cyber Threat Intelligence}
\newacronym{llm}{LLM}{Large Language Model}
\newacronym{misp}{MISP}{Malware Information Sharing Platform}
\newacronym{shacl}{SHACL}{Shapes Constraint Language}
\newacronym{stix}{STIX}{Structured Threat Information Expression}
\newacronym{ttp}{TTP}{Tactics, Techniques, and Procedures}
\newacronym{uco}{UCO}{Unified Cyber Ontology}
\newacronym{kg}{KG}{Knowledge Graph}
\newacronym{rag}{RAG}{Retrieval Augmented Generation}
\newacronym{rq}{RQ}{Research Questions}
\newacronym{mmr}{MMR}{Maximal Marginal Relevance}
\newacronym{rdf}{RDF}{Resource Description Framework}

%% file: parts/abstract.tex
\begin{abstract}
Effective \gls{cti} relies upon accurately structured and semantically enriched information extracted from cybersecurity system logs. However, current methodologies often struggle to identify and interpret malicious events reliably and transparently, particularly in cases involving unstructured or ambiguous log entries. In this work, we propose a novel methodology that combines ontology-driven structured outputs with \glspl{llm}, to build an \gls{ai} agent that improves the accuracy and explainability of information extraction from cybersecurity logs. Central to our approach is the integration of domain ontologies and SHACL-based constraints to guide the language model's output structure and enforce semantic validity over the resulting graph. Extracted information is organized into an ontology-enriched graph database, enabling future semantic analysis and querying. The design of our methodology is motivated by the analytical requirements associated with honeypot log data, which typically comprises predominantly malicious activity. While our case study illustrates the relevance of this scenario, the experimental evaluation is conducted using publicly available datasets. Results demonstrate that our method achieves higher accuracy in information extraction compared to traditional prompt-only approaches, with a deliberate focus on extraction quality rather than processing speed.
\end{abstract}

%% file: parts/introduction.tex
The increasing complexity of cyber threats has led to a greater reliance on intelligence-oriented approaches within cybersecurity~\cite{hutchins2011intelligence, mcdaniel2016towards, tounsi2018survey}. \gls{cti} contributes to this development by converting unstructured data into structured forms that support threat detection, analysis, and response (\Cref{sec:background}). Among available data sources, system logs---particularly those collected from honeypot or honeynet deployments---are especially relevant due to their significant concentration of adversarial activity~\cite{nawrocki2016survey}. Unlike routine operational environments, which generate logs dominated by benign entries, honeypots are designed to attract malicious interactions, making the resulting data more relevant for analysis.

Log data remains challenging to process. It is typically unstructured, syntactically inconsistent, and often ambiguous in meaning, which complicates automated interpretation. Traditional techniques based on fixed rules or heuristics tend to be limited in adaptability and are sensitive to variation in threat behavior. More recent methods involving \glspl{llm} have shown improved capabilities in extracting information from natural language~\cite{izacard2020leveraging, lewis2020retrieval, srivastava2022beyond}. However, they remain general in scope and can lack precision when applied to domain-specific content.

This work outlines a methodology (\Cref{sec:methodology}), named "OntoLogX", aimed at facilitating more transparent extraction of \gls{cti} from logs. The approach incorporates prompt-driven interaction with \glspl{llm}, guided by a domain-specific ontology designed to represent the structural and contextual characteristics of cybersecurity-relevant data. The extracted information is structured into a \gls{kg}, enabling semantic querying, traceability, and integration with \gls{cti} workflows. The use of natural language interfaces may contribute to greater transparency in processing, as the logic of data extraction can be expressed and modified through prompts. The ontology also supports the organization of extracted information into knowledge graphs and enables the enforcement of quality constraints through \gls{shacl}-based validation.

We demonstrate that the proposed approach significantly improves information extraction accuracy over a traditional, prompt-only baseline method (\Cref{sec:experiment}). While the methodological design of this research draws on the characteristics of honeypot data, the experimental evaluation is conducted using publicly available log datasets to ensure reproducibility and comparability. Extracted intelligence is stored in an ontology-enriched graph database, enabling future semantic querying and downstream \gls{cti} tasks.

We conclude the paper by outlining directions for future work (\Cref{sec:conclusions}). While the present methodology demonstrates accuracy in information extraction from cybersecurity logs, further development is required to ensure operational relevance for \gls{cti} analysis.

%% file: parts/background.tex
\subsection{The Need for \glsentrylong{cti}}

The escalation in complexity, sophistication, and frequency of cyber threats has increasingly exposed the limitations of traditional reactive cybersecurity strategies, necessitating a shift towards proactive and anticipatory methods~\cite{hutchins2011intelligence, mcdaniel2016towards}. \Gls{cti} addresses this need by systematically collecting, transforming, and analyzing vast amounts of threat-related data, transforming it into actionable knowledge capable of supporting informed decision-making processes within cybersecurity operations~\cite{tounsi2018survey}. Effective \gls{cti} facilitates not only threat detection and mitigation but also proactive threat hunting and risk management, thus significantly enhancing organizational cyber resilience~\cite{wagner2019cyber}.

A critical source of \gls{cti} derives from honeypots or honeynets---deliberately vulnerable systems explicitly designed to attract and capture malicious cyber activity. Logs produced by these systems inherently record interactions from malicious actors, offering a unique dataset rich with insights into attacker behavior, tactics, techniques, and emerging threat trends~\cite{spitzner2003honeypots}. Unlike traditional security logs, which are often overloaded with benign or irrelevant entries, honeypot-generated logs predominantly represent genuine threats, making them particularly valuable for focused, accurate intelligence extraction and threat analysis~\cite{nawrocki2016survey}.

However, despite their value, honeypot logs pose significant analytical challenges due to their inherently heterogeneous, unstructured, and semantically ambiguous nature~\cite{fraunholz2018investigation}. Traditional heuristic-based or pattern-matching approaches struggle to reliably extract comprehensive and accurate information from these logs, particularly when dealing with sophisticated attacks exhibiting novel or evolving characteristics. Consequently, there is an explicit requirement for more advanced methods capable of interpreting and structuring these logs accurately and semantically, ensuring the reliability and efficacy of derived \gls{cti}.

This challenge underscores the critical importance of innovative methodologies, such as \glspl{llm} combined with semantic technologies~\cite{zhao2024ontology}. The integration of ontology-enriched representations into \gls{cti} workflows significantly enhances semantic clarity and precision, enabling more accurate interpretation, classification, and contextualization of threats derived from honeypot logs. Leveraging such structured semantic frameworks ultimately improves the overall quality and actionable value of threat intelligence, substantially reinforcing cybersecurity preparedness and response~\cite{syed2016uco, barnum2012standardizing}.

\subsection{Fundamental Concepts of Ontologies and Knowledge Graphs}

Ontologies are commonly defined as formal, explicit specifications of shared conceptualizations encompassing classes, relationships, and constraints within a given domain~\cite{studer1998knowledge}. In cybersecurity, ontology-based frameworks are increasingly adopted to organize and standardize threat-related knowledge, enabling semantic interoperability, automated reasoning, and improved information integration across heterogeneous sources~\cite{syed2016uco}.
 
Many widely adopted cybersecurity ontologies are available in the literature, each addressing specific needs within \gls{cti} analysis. \gls{uco}~\cite{syed2016uco} provides comprehensive concepts and relationships for broad cybersecurity knowledge integration; \gls{stix}~\cite{barnum2012standardizing} defines a widely adopted standard for \gls{cti} exchange; CRATELO~\cite{oltramari2014building} specifically targets the representation of cyber incidents and forensic data; \gls{misp}~\cite{wagner2016misp} facilitates structured malware and threat indicator sharing. Particularly relevant to our use case is the SEPSES ontology~\cite{KieslingEKE19}, which provides a rich, structured vocabulary for the semantic integration of already-parsed logs, but is not intended to guide information extraction from free-text messages or support interaction with language models.

The use case of real-time extraction from log messages using language models, however, benefits from an ontology that is minimal, interpretable, and structurally constrained for validation. Ontologies that rely on preprocessed observables, normalized fields, or syntactically consistent event abstractions, do not adapt easily to this task. To address this, we introduce a lightweight ontology designed for structured output parsing and SHACL-conformant triple generation. While it does not aim for the expressive depth of standard cybersecurity ontologies, it serves as an intermediate semantic layer suitable for direct anchoring in \gls{llm} workflows and subsequent alignment with established models.

A \gls{kg} can be defined as a graph of data intended to accumulate and convey knowledge of the real world, whose nodes represent entities of interest and whose edges represent potentially different relations between these entities~\cite{hogan2021knowledge}. In practice, \glspl{kg} are typically grounded in an ontology that ensures semantic consistency and facilitates reasoning, validation, and integration of heterogeneous data sources.

The quality and ontology-compliance of \glspl{kg} can be verified through the use of \gls{shacl}~\cite{knublauch2017shapes} constraints. \gls{shacl} provides a declarative framework to define and validate structural and semantic constraints. When integrated with ontology-based representations, \gls{shacl} enables automated quality control over knowledge, verifying whether generated \glspl{kg} respect required types, property cardinalities, and relationship patterns. This validation process helps detect inconsistencies, enforce domain rules, and ensure that the resulting \glspl{kg} remain semantically coherent and human-auditable, even when derived from noisy or inferred inputs. In the context of log analysis, \gls{shacl} plays a vital role in guaranteeing that automatically generated semantic representations conform to expected schema designs, thereby enhancing both the robustness and explainability of downstream \gls{cti} tasks~\cite{pareti2022knowledge, RabbaniLH23}.

\subsection{Large Language Models, Retrieval Augmented Generation, and AI Agents}

% Consequently, there is an explicit need to integrate \glspl{llm} within ontology-driven pipelines that leverage structured semantic knowledge to guide the reasoning and extraction processes~\cite{jiang2023knowledge,sharma2024}. Ontologies offer formal representations of domain-specific concepts and relationships, significantly enhancing \gls{llm} performance by providing semantic contexts and constraints to guide inference and interpretation. In cybersecurity, such semantic guidance is crucial, as accurate identification and categorisation of complex threat-related information directly influence the quality of subsequent intelligence analysis.

% In addition to ontology-driven semantic enhancement, it is essential to incorporate explicit data-quality constraints to ensure that information extracted by \gls{llm} pipelines is reliable and actionable. 

% Existing research has demonstrated that combining semantic constraints with agentic language model frameworks leads to substantially improved accuracy and reliability of extracted information~\cite{sun2023ontology}. Hence, the explicit integration of ontology-driven semantic knowledge and \gls{shacl}-based constraints within agentic \gls{llm} pipelines constitutes a promising strategy for developing robust and semantically precise cybersecurity intelligence extraction systems. Such integration not only facilitates the reuse of existing ontological knowledge but also explicitly enforces data quality standards, thereby enhancing the practical utility and dependability of \gls{cti} derived from complex log datasets.

\Glspl{llm} are transformer-based models trained on extensive textual corpora to perform a broad range of natural language understanding and generation tasks~\cite{vaswani2017attention, devlin2019bert, brown2020language}. These models learn probabilistic representations of language, enabling them to complete, summarize, translate, and interpret text across multiple domains. Their performance has been demonstrated across benchmarks in zero-shot and few-shot learning~\cite{srivastava2022beyond}. However, despite their versatility, \glspl{llm} do not inherently guarantee factual consistency, structural coherence, or domain-specific accuracy. Outputs may reflect biases in the training data or lack sufficient grounding in external knowledge, particularly when applied to specialized domains such as cybersecurity, where terminology and contextual interpretation are often critical.

One strategy to address these limitations is \Gls{rag}, which combines language generation with information retrieval to enhance factual grounding~\cite{lewis2020retrieval}. In the \gls{rag} framework, a retriever component selects documents relevant to a given query from a pre-indexed knowledge base. The retrieved content is then passed to a language model, which conditions its output on this external evidence. This method allows for more contextually informed and domain-relevant responses, especially in scenarios where training data alone may be insufficient or outdated~\cite{izacard2020leveraging, guu2020retrieval}. In the context of cybersecurity, it offers the potential to improve information extraction from unstructured sources such as system logs, where background knowledge may be necessary to interpret ambiguous or incomplete entries.

While the integration of \glspl{llm} with retrieval mechanisms improves output relevance, challenges remain with respect to consistency, interpretability, and quality control. These concerns are particularly salient in applications involving automated extraction of threat indicators or behavioral patterns from log data, where the reliability of each extracted item has downstream implications for analysis and response. In such cases, the transparency of the processing pipeline---understood here as the ability to inspect and guide the model’s decision process through natural language prompts and accessible intermediate outputs---becomes an important consideration. Transparent configurations facilitate manual oversight and error correction, and may also support reproducibility in operational settings.

AI Agents can be defined as autonomous software entities, typically based on \glspl{llm}, engineered for goal-directed task execution within bounded digital environments~\cite{acharya2025agentic, sapkotaAIAgentsVs2025a, sadoExplainableGoaldrivenAgents2023}. These agents can perceive structured or unstructured inputs, reason over contextual information, initiate actions toward achieving specific objectives, often acting as surrogates for human users or subsystems~\cite{sapkotaAIAgentsVs2025a}.

Existing research has demonstrated that combining semantic constraints leads to substantially improved accuracy and reliability of extracted information~\cite{sun2023ontology}. Hence, the explicit integration of ontology-driven semantic knowledge and \gls{shacl}-based constraints within agentic \gls{llm} pipelines constitutes a promising strategy for developing robust and semantically precise cybersecurity intelligence extraction systems.

%% file: parts/methodology.tex
OntoLogX operates as an \gls{ai} Agent with minimal user interaction, constructing a \gls{kg} representing a log event and, optionally, its contextual information. It integrates an \gls{llm} with an ontology for log events to leverage symbolic representations of the cybersecurity domain, thereby enabling automatic feedback behaviors that ensure the generation of a valid \gls{kg}. OntoLogX is designed for online operation, parsing and processing log events incrementally and sequentially, one by one, in a setting that reflects realistic cybersecurity use cases requiring near real-time event analysis.

\begin{figure}[tb!]
    \centering
    \includegraphics[width=.8\linewidth]{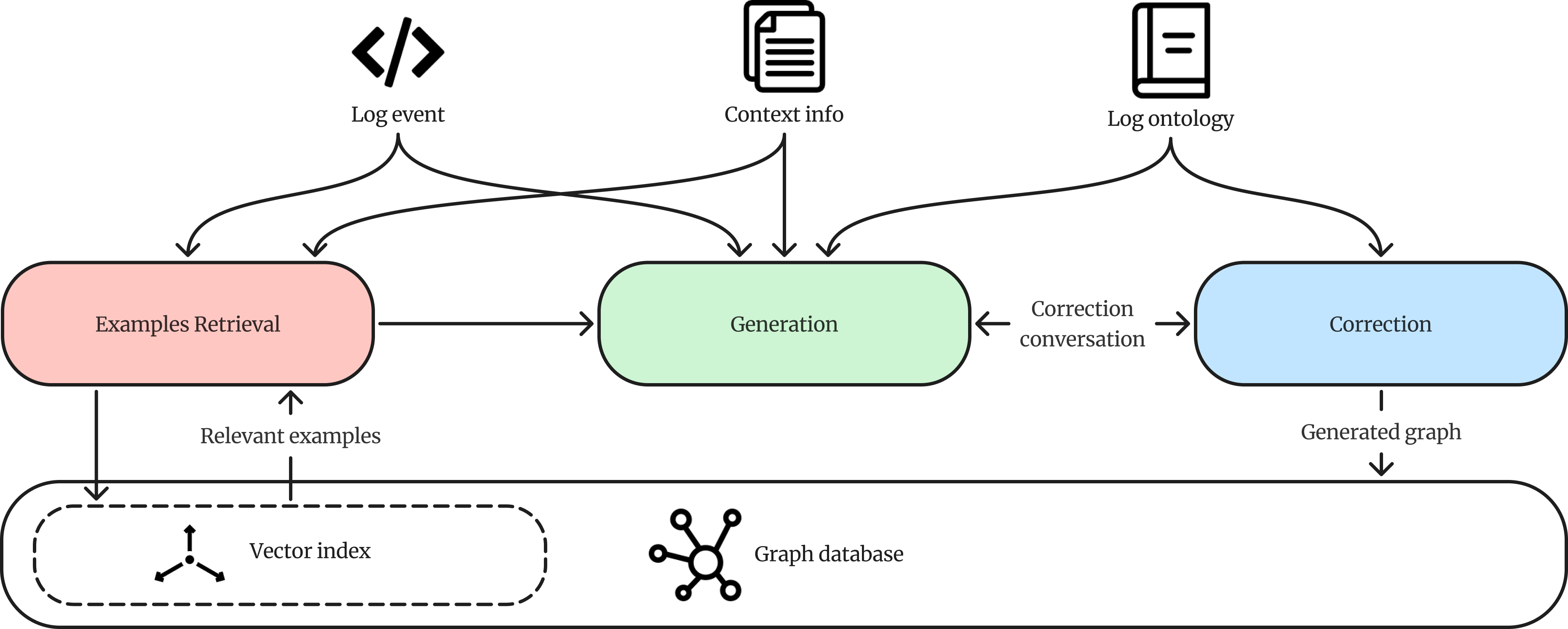}
    \caption{High-level procedure for generating a log event \gls{kg}.}
    \label{fig:methodology}
\end{figure}

\Cref{fig:methodology} depicts the overall workflow. Given a log event and any optional context information (e.g., the device or application the log originates from), several relevant log event \glspl{kg} are retrieved from the graph database. Then, the input log event, context information, and the log ontology are used to instruct an \gls{llm} to generate a new \gls{kg}, using the retrieved \glspl{kg} as few-shot examples. The generated \gls{kg} is validated: if it does not conform to the expected format or is not ontology-compliant, the \gls{llm} is prompted again in the same conversation to generate specific corrections. If a valid \gls{kg} is eventually produced, it is saved in the graph database where it may be retrieved during future log processing operations. It is worth noting that \glspl{kg} are stored independently from each other, as the semantic connection of different \glspl{kg} is not the focus of this work.

\subsection{Log Ontology}\label{sec:log-ontology}

\begin{figure}[htb]
    \centering
    \includegraphics[width=.8\linewidth]{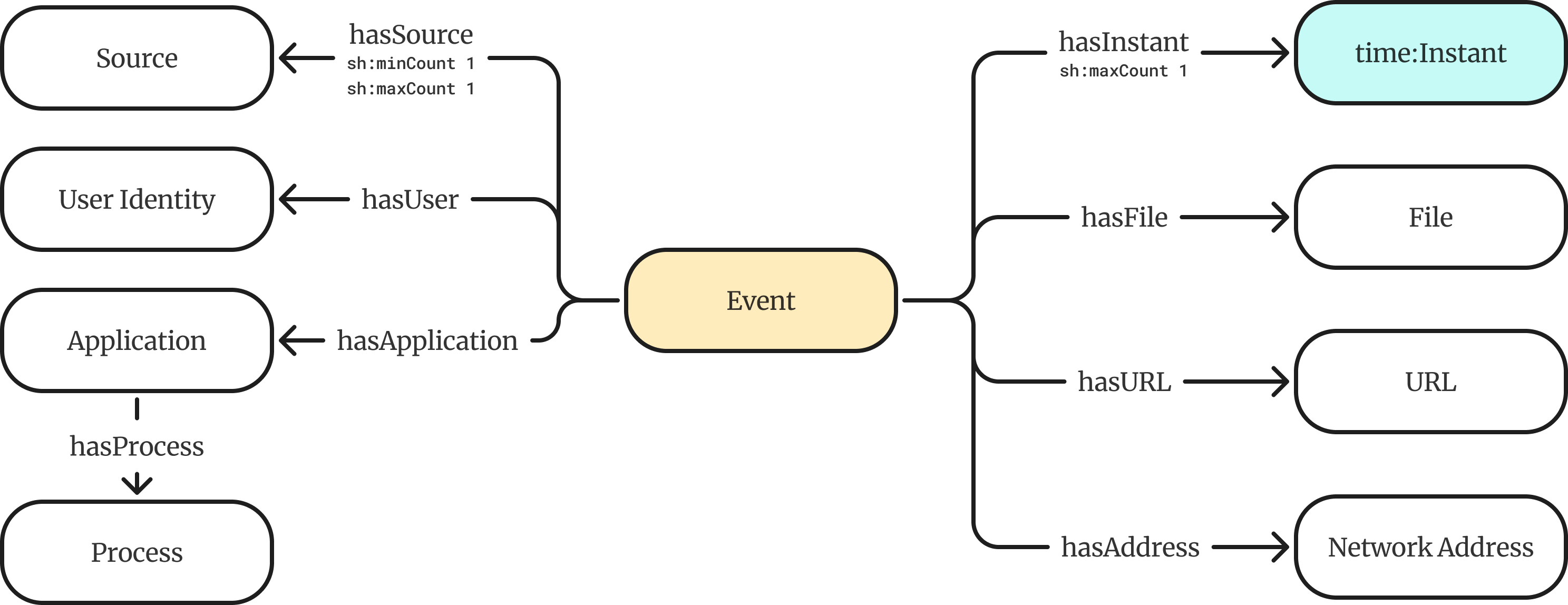}
    \caption{Representation of the classes and object properties of the OntoLogX ontology, with SHACL constraints that apply to them. Data properties are not reported for conciseness.}
    \label{fig:ontology}
\end{figure}

To formalize the knowledge to be inferred from log events and enable semantic interoperability of the generated \glspl{kg}, we developed a custom log ontology, illustrated in \Cref{fig:ontology}. Starting from the recommended log event details by the Australian Signals Directorate’s Australian Cyber Security Centre\footurl{https://www.cyber.gov.au/sites/default/files/2024-08/best-practices-for-event-logging-and-threat-detection.pdf}, we adopted a bottom-up, data-driven approach to design a lightweight, flat ontology grounded in the direct analysis of representative log datasets, including but not limited to honeynet event streams. While it does not aim for full expressivity, it provides a stable intermediate representation suitable for post-processing alignment with richer ontologies such as STIX or SEPSES. For example, a log entry such as:
\[
\texttt{2022-01-21 03:49:44 jhall/192.168.230.165:46011 VERIFY OK: CN=OpenVPN CA}
\]
implicitly contains information such as a timestamp, user identity, network address, and certificate details—all encoded without structural consistency or explicit separation. Ontologies such as \gls{uco} or \gls{stix} typically assume the availability of structured metadata or pre-parsed fields, making them difficult to apply directly to such unstructured inputs.

\input{media/tables/ontology-comparison}

\Cref{tab:ontology-comparison} summarizes the key differences between our ontology and widely adopted cybersecurity schemas in terms of design goals, input assumptions, and applicability to log-based semantic extraction.

The primary design objective was to create a self-contained, semantically expressive model that captures the core entities and relationships found in raw log entries without being too specific for our specific use case, thus making the ontology scalable to logs from different datasets and sources. The central concept is the class \texttt{Event}, which serves as the semantic anchor for each log entry. Associated classes are also modeled: \texttt{User Identity}, the credential or identity of a user involved in the event; \texttt{Application} and \texttt{Process}, used to represent execution context; \textit{File}, the file resource being accessed or modified; \texttt{Network Address}, such as IP addresses or ports mentioned in the log; \texttt{Source}, the logical or physical origin of the log entry; \texttt{URL}: representing any remote reference in the log content. Temporal dimensions are modeled using the standard W3C Time ontology\footurl{https://www.w3.org/TR/owl-time/}. In addition to object properties (e.g., \texttt{hasUser}, \texttt{hasProcess}), the ontology includes data properties such as \texttt{eventMessage}, capturing the log’s raw content, and \texttt{logLevel}, denoting its severity.

To enforce schema compliance and semantic consistency, a companion SHACL specification defines constraints on key classes and properties. These constraints are critical when constructing graphs via a \gls{llm}-based extraction pipeline, where enforcing schema compliance helps mitigate issues such as missing fields or incorrect relationships. By enforcing constraints on cardinality, type consistency, and presence of essential properties, we ensure that generated \glspl{kg} not only follow the OntoLogX vocabulary but are also semantically valid and queryable. 

\subsection{Examples Retrieval}\label{sec:examples-retrieval}

To guide the generation process, the raw log event, along with optional context (e.g., the device or application that generated the log), is first queried against a vector store. This store indexes both previously generated \glspl{kg} and manually crafted examples aligned with the log ontology. The goal of this retrieval step is to identify semantically similar examples that can serve as few-shot prompts, thereby enhancing the quality and structure of the output.

The search leverages \gls{mmr}~\cite{carbonell1998use}, a ranking strategy designed to balance \textit{relevance} and \textit{diversity}. Rather than simply returning the top-$k$ most similar entries, \gls{mmr} penalizes redundancy by selecting items that are both relevant to the query and dissimilar from one another. This reduces the likelihood of retrieving near-duplicate examples and encourages exposure to a broader range of graph structures and content patterns. Formally, given a query $q$, a candidate set of documents $D$, and a set of already selected items $S \subset D$, \gls{mmr} iteratively selects the next item $d^* \in D \setminus S$ that maximizes the following objective:
\[
\text{MMR}(d) = \lambda \cdot \text{Sim}(d, q) - (1 - \lambda) \cdot \max_{s \in S} \text{Sim}(d, s)
\]
where $\text{Sim}(d, q)$ measures the similarity between document $d$ and query $q$, $\text{Sim}(d, s)$ measures the similarity between $d$ and already selected document $s$, and $\lambda \in [0,1]$ controls the trade-off between relevance and diversity. This retrieval strategy helps the \gls{llm} better internalize the expected structure and semantics of the output graph. Additionally, it enables the model to reuse domain knowledge and contextual cues from previously processed log events, improving both consistency and completeness in \gls{kg} construction.

\subsection{Generation}

The generation component of OntoLogX is responsible for constructing a \gls{kg} from raw log events in alignment with the log ontology described in \Cref{sec:log-ontology}. This step involves invoking an \gls{llm}, which takes as input the raw log entry, any optional contextual information, and a set of few-shot examples retrieved through semantic search. The use of a \gls{llm} enables the generation of high-quality \glspl{kg} even in low-resource scenarios, without requiring task-specific supervised training. Furthermore, the \gls{llm} can leverage its internal knowledge to infer implicit information, resolve ambiguities, and normalize terminology—capabilities that are often difficult to replicate with rule-based or statistical parsers.

We hypothesize that \glspl{llm} inherently encode significant amounts of latent cybersecurity knowledge, due to their pretraining on diverse and large-scale datasets. This domain familiarity improves their ability to recognize entities, activities, and relationships common in log data, even when such elements are expressed in non-standard, abbreviated, or noisy formats. Consequently, an \gls{llm}-driven approach offers significantly greater adaptability to heterogeneous or previously unseen log types compared to conventional log parsers, which typically depend on learned templates to extract parameters from structured inputs.

\input{media/listings/main-prompt}

The prompt used to instruct the \gls{llm} (shown in \Cref{lst:main-prompt}) is intentionally designed to be model-agnostic, maximizing compatibility with a broad range of language models. It comprises: (i) a clear role and task definition, (ii) detailed instructions grounded in the OntoLogX ontology, and (iii) a set of constraints and clarifications based on common failure patterns observed in early experiments, such as malformed URIs, incorrect predicate usage, and mismatched types.

\input{media/tables/structured-output}

To reinforce structural consistency and reduce formatting variability, the \gls{llm} is equipped with a structured output tool, whose target schema is shown in \Cref{tab:structured-output}. This tool defines the required fields and their expected types, effectively serving as a strong constraint during generation. It not only improves the reliability of downstream parsing and validation but also helps encode explicit knowledge about the ontology, ensuring that the produced \glspl{kg} conform to the intended semantic structure.

Finally, the \gls{llm} is provided with few-shot examples---using the \glspl{kg} retrieved from the database in the previous step---further reinforcing the desired output structure and ontology-aligned vocabulary, guiding the model toward generating consistent and valid outputs.

\subsection{Correction}

Given the inherent variability of \glspl{llm}, the output of the generation step may fail to conform to the expected ontology structure, or it may be incomplete or semantically invalid. To address this, OntoLogX incorporates a dedicated \textit{correction} phase that identifies such issues and engages in a feedback-driven interaction with the \gls{llm} to request targeted revisions.

This correction process is coordinated through the \gls{llm}'s structured output tool, which performs automatic validation using two key criteria. First, it assesses the syntactic validity of the output, verifying that the graph is composed of well-formed nodes and relationships. Second, it checks for ontology compliance, ensuring that the graph adheres to the structural and semantic constraints specified by the log ontology—this includes correct class and property usage, proper data typing, and satisfaction of required schema constraints.

When one or more issues are detected, a tailored correction prompt is constructed and passed back to the \gls{llm}. This iterative interaction continues for multiple rounds until a fully valid and ontology-compliant \gls{kg} is produced. If a graph without errors \gls{kg} is not achieved within these rounds, an empty graph is instead produced, thus penalizing evaluation metrics.

Once a syntactically valid and ontology-compliant \gls{kg} is produced---either on the first attempt or after one or more correction cycles---it is stored in the graph database for long-term persistence and downstream consumption. Each \gls{kg} is also annotated with its input log event and contextual information, to enable traceability and further querying. This source information is also used to compute a vector embedding of each \gls{kg}, allowing for efficient semantic search and retrieval of related events in future queries.

%% file: media/tables/ontology-comparison.tex
\begin{table}
\caption{Comparison of the suitability of selected cybersecurity ontologies for log-based semantic extraction.}
\label{tab:ontology-comparison}
\centering
\renewcommand{\arraystretch}{1.2}
\resizebox{\textwidth}{!}{%
\begin{tabular}{@{}lllll@{}}
\toprule
Aspect & OntoLogX ontology & SEPSES ontology & UCO & STIX \\
\midrule
Purpose & LLM-based log parsing & Log integration & Unified CTI modeling & Threat intel exchange \\
Input Type & Free-text logs & Parsed logs & Structured observables & Structured indicators \\
Complexity & Minimal, flat schema & Medium-high & High, modular & High, domain-specific \\
SHACL Use & Core for validation & RDF quality checks & Optional & No \\
Ontology Role & Preprocessing layer & Integration layer & Reasoning layer & Exchange format \\
Mapping Potential & Core CTI entities & CVE, CWE, CAPEC, CPE & STIX, CVE, CAPEC & Often paired with UCO \\
\bottomrule
\end{tabular}
}
\end{table}

%% file: media/listings/main-prompt.tex
\begin{table}
\caption{Prompt for log event \gls{kg} generation, used in conjunction with structured output.}
\label{lst:main-prompt}
\begin{minted}
[
frame=lines,
framesep=2mm,
baselinestretch=1,
fontsize=\footnotesize,
breaklines=true,
breaksymbolleft={},
breaksymbolright={}
]
{markdown}
# Overview
You are a top-tier cybersecurity expert specialized in extracting structured information from unstructured data to construct a knowledge graph according to a predefined ontology. You will be provided with a log event, optionally accompanied by contextual information.
Your goal is to maximize information extraction from the event while maintaining absolute accuracy. Leverage both the contextual information and your knowledge of computer systems and cybersecurity to infer additional insights where possible. The objective is to achieve completeness in the knowledge graph while remaining strictly ontology-compliant.
# Rules
You MUST adhere to the following constraints at all times:
- The graph must contain exactly one "Event" node, with a property "eventMessage" that holds the original event text.
- Do not introduce any new node types, relationship types, or property types. Only use the available types.
- Respect the appropriate casing for all types.
- Use the appropriate node prefix for properties, e.g. "userUID" instead of "uid".
- If not specified in the log event, try to infer the severity of the event based on the message. If you cannot determine the severity, default to "INFO".
- The graph must be connected: there should be no isolated nodes.
# Strict Compliance
Adhere to these rules strictly. Any deviation will result in termination.
\end{minted}
\end{table}

%% file: media/tables/structured-output.tex
\begin{table}
\caption{Format of structured output that the \gls{llm} must conform to.}
\label{tab:structured-output}
\centering
\resizebox{\textwidth}{!}{%
\renewcommand{\arraystretch}{1.2}
\begin{tabular}{
@{}
l
>{\raggedright\arraybackslash}m{0.4\textwidth}  % Description
l
>{\raggedright\arraybackslash}m{0.35\textwidth}  % Element Description
@{}
}
\toprule
\multirow{2}{*}{Entity} & \multirow{2}{*}{Description} & \multicolumn{2}{l}{Elements} \\ \cmidrule(l){3-4} 
 &  & Name & Description \\ \midrule
\multirow{2}{*}{Graph} & \multirow{2}{=}{An event knowledge graph composed of nodes and relationships.} & Nodes & List of nodes in the graph. \\
 &  & Relationships & List of relationships in the graph. \\ \hline
\multirow{3}{*}{Node} & \multirow{3}{=}{A node in the event graph. Each node type has a specific set of allowed properties.} & ID & Unique identifier for the node. \\
 &  & Type & Ontology class name. \\
 &  & Properties & List of properties of the node. \\ \hline
\multirow{2}{*}{Property} & \multirow{2}{=}{A property of a node in the event knowledge graph.} & Type & Ontology data property name. \\
 &  & Value & Extracted value of the property. \\ \hline
\multirow{3}{*}{Relationship} & \multirow{3}{=}{A relationship between two nodes in the event graph. Each relationship type has a predefined source and target node type.} & Source ID & Unique identifier of source node. \\
 &  & Target ID & Unique identifier of target node. \\
 &  & Type & Ontology object property name. \\ \bottomrule
\end{tabular}%
}
\end{table}

%% file: parts/experiments.tex
\input{media/listings/baseline-prompt}

This section compares the experimental results obtained using OntoLogX and a baseline configuration designed for comparative evaluation. The baseline consists of a single generation step based on a traditional prompt-only approach, thus excluding the examples retrieval and correction features of OntoLogX. The baseline prompt, which is obtained by adding the lines shown in \Cref{lst:baseline-prompt} to the prompt in \Cref{sec:examples-retrieval}, instructs the \gls{llm} in producing a JSON-formatted output conforming to the schema in \Cref{tab:structured-output} with plain prompting rather than with structured output.

\input{media/tables/llms}

Both methodologies were evaluated across a total of five different \glspl{llm}, which are reported in \Cref{tab:llms}. Notably, the proprietary models chosen are significantly more expensive than those with more permissive licenses, as are medium-sized models compared to smaller ones.

All models were accessed via AWS Bedrock\footurl{https://aws.amazon.com/it/bedrock/}, except for \texttt{Qwen 2.5 Coder}, which was run locally using vLLM~\cite{kwon2023efficientmemorymanagementlarge} on a system equipped with four NVIDIA T4 GPUs. A temperature of 0.7 was used over all runs to promote creative reasoning, which we hypothesize is beneficial for inferring implicit information from raw logs. To mitigate the variability introduced by this setting, each experiment was repeated ten times.

\subsection{Implementation}

OntoLogX is implemented in Python, using the LangChain framework\footurl{https://www.langchain.com/} to manage the \gls{llm} and parsing pipeline. The resulting ontology-compliant \glspl{kg}, along with the underlying log ontology, are stored in a Neo4j graph database\footurl{https://neo4j.com/product/neo4j-graph-database/}, extended with a vector index to support efficient storage and semantic retrieval based on node property embeddings. To facilitate reproducibility and the structured sharing of results, the database is organized with a graph schema that follows the MLSchema ontology~\cite{publio2018MLSchema}.

The structured output format is specified through the Pydantic\footurl{https://pydantic.dev/} library, and the \glspl{llm} generate the output through function calling.

In the version evaluated in this study, only manually annotated examples are used for few-shot prompting; graphs previously generated by the system are excluded from the retrieval step. The embedding model used is \texttt{gte-multilingual-base}. Moreover, the correction step is limited to a maximum of three refinement prompts per log. If a valid graph is not obtained within this limit, the output is considered to be an empty graph.

\subsubsection{Dataset}

The test dataset consists of log events sampled from the AIT-LDS dataset~\cite{landauerAITLogData2022}. A total of 70 log entries were randomly selected to ensure broad diversity, both in log types (e.g., Apache, VPN, audit) and semantic content. To promote this diversity, the first 100 log events were extracted from each file in the \textit{RussellMitchell} testbed. From this pool, 70 events were iteratively selected using an embedding-based dissimilarity criterion.

Specifically, the embedding of each candidate log entry was computed using the \texttt{nomic-embed-text-v1.5} model. For each newly drawn candidate, the cosine distance was calculated with respect to the embeddings of all previously selected entries. If the minimum distance was below 0.7—indicating excessive semantic similarity—the candidate was retained; otherwise, it was discarded and another was drawn. Each selected log entry was manually annotated with a corresponding ground-truth \gls{kg}, and the dataset was partitioned into three subsets: the first 10 events were reserved for few-shot prompting, the next 10 for validation, and the remaining 50 for testing.

\subsubsection{Metrics}\label{sec:metrics}

To evaluate the quality of \glspl{kg} generated by \glspl{llm}, we combine the precision, recall and F1 score metrics from the ontologies field with G-Eval \glspl{llm}. In addition to these metrics, we evaluate the percentage of generated graphs that violate SHACL rules.
\begin{enumerate}
    \item \textit{Precision}: ratio of generated triples that are correct, i.e., have a matching triple in the ground truth, over the total number of generated triples. High precision values indicate that the model tends to produce accurate triples with relatively few incorrect or spurious facts.
    \item \textit{Recall}: ratio of correct triples that were generated, i.e., those that match a ground-truth triple, over the total number of ground-truth triples. High recall values indicate that the model successfully captures most of the relevant information present in the ground truth.
    \item \textit{F1 Score}: harmonic mean between precision and recall.
    \item \textit{G-Eval Score}: a \gls{llm}-as-a-judge framework that uses chain-of-thought reasoning and a form-filling paradigm to evaluate natural language generation outputs~\cite{liu2023gevalnlgevaluationusing}. It operates by prompting the \gls{llm} with a set of evaluation criteria---either user-defined or automatically derived---to produce scores reflecting various aspects of output quality. In our setting, it uses the \texttt{Llama 3.3} model to assess the extent to which the generated \gls{kg} retains the original log event's meaning. The \gls{llm} is prompted to produce natural language summaries of both the raw log and the \gls{kg}, and then to score their semantic overlap on a scale from 0 to 1. Additional information in the \gls{kg} lowers the score only if it is judged irrelevant. Higher G-Eval scores indicate more faithful and relevant extractions.
\end{enumerate}

\subsubsection{Results}

\input{media/tables/model-performance}

\begin{figure}
    \begin{subfigure}{.5\textwidth}
        \centering
        \includegraphics[width=.9\linewidth]{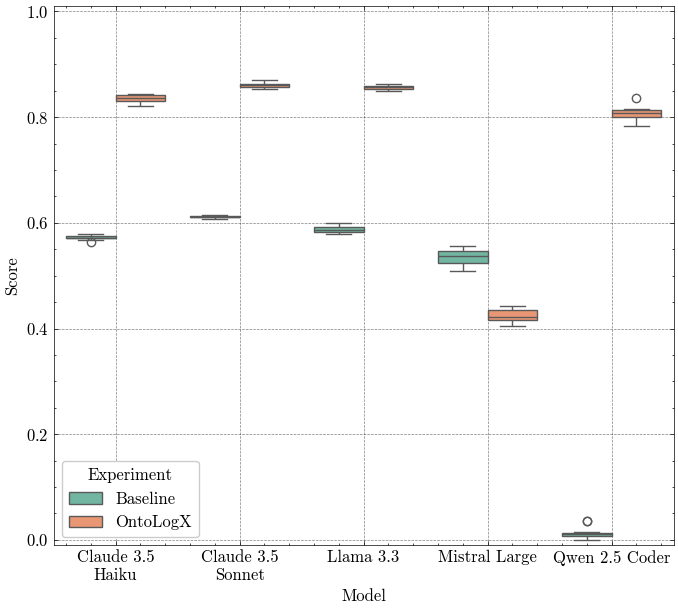}
        \caption{F1 scores}%
        \label{fig:f1}
    \end{subfigure}%
    \begin{subfigure}{.5\textwidth}
        \centering
        \includegraphics[width=.9\linewidth]{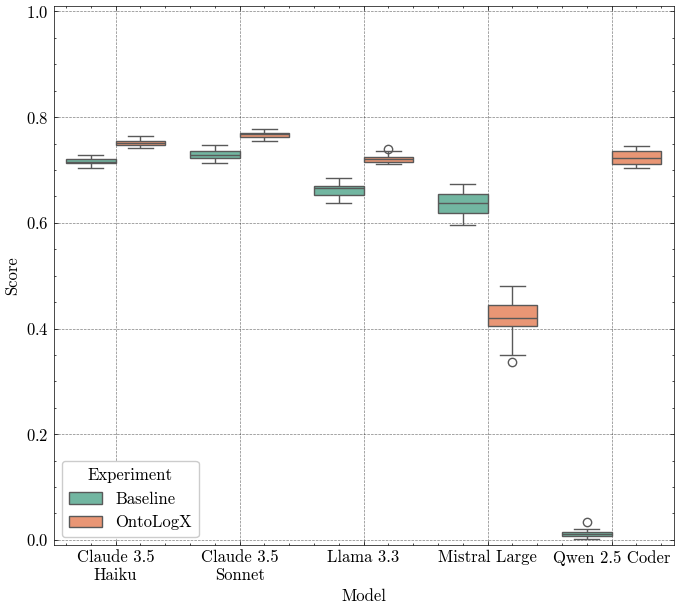}
        \caption{G-Eval scores}%
        \label{fig:g-eval}
    \end{subfigure}
    \begin{subfigure}{\textwidth}
        \centering
        \includegraphics[width=.45\linewidth]{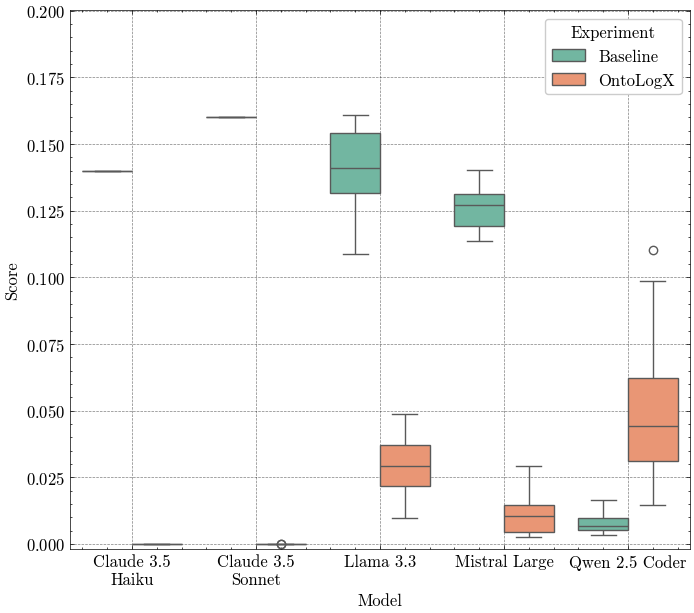}
        \caption{Percentage of graphs that violate SHACL constraints}
        \label{fig:shacl-violations}
    \end{subfigure}
    \caption{Comparison of selected metrics between baseline and OntoLogX}
  \label{fig:scores}
\end{figure}

\Cref{tab:model-performance} reports the mean and standard deviation over all runs of each model under the baseline and OntoLogX configurations. The mean values for the F1 and G-Eval scores, along with the percentage of SHACL violations, are also reported in \Cref{fig:scores}. Overall, the execution time---encompassing the examples retrieval, generation, and corrections but excluding database storage time---goes from 1.4 seconds with \texttt{Llama 3.3} to 9.8 seconds with \texttt{Mistral Large}.

Across nearly all models and metrics, the OntoLogX-enhanced setup consistently outperforms the baseline configuration. In particular:
\begin{itemize}
\item All models except \texttt{Mistral Large} achieve higher precision under the OntoLogX setup. The most significant improvement is observed with \texttt{Qwen 2.5 Coder}, for which the baseline configuration yields highly variable results.
\item OntoLogX substantially improves recall for all models except \texttt{Mistral Large}, which shows a decrease. \texttt{Claude 3.5 Sonnet} and \texttt{Qwen 2.5 Coder} exhibit the largest gains.
\item OntoLogX improves G-Eval scores across all models. Again, \texttt{Qwen 2.5 Coder} benefits the most, suggesting enhanced semantic alignment and structural quality in its outputs.
\item OntoLogX also significantly increases SHACL compliance, mostly due to the use of structured outputs that conform to the SHACL rules. Interestingly, the results are not guaranteed to be fully SHACL-compliant: this can be attributed to the limited expressiveness offered by the adopted structured output approach compared to the SHACL constraints language.
\end{itemize}

Among all evaluated models, \texttt{Claude 3.5 Sonnet} stands out as the top performer in the OntoLogX configuration, achieving the highest F1 and G-Eval scores, alongside strong results in the other metrics. Nevertheless, \texttt{Llama 3.3} emerges as a robust open-weights alternative with competitive performance. Conversely, \texttt{Mistral Large} is the only model that performs better under the baseline configuration for recall, F1, and G-Eval—though the absolute values remain modest. This outcome suggests a lower compatibility with the OntoLogX pipeline or a higher sensitivity to structured prompting and tool-assisted outputs. It also underscores the limitations of adopting a uniform prompting strategy across heterogeneous models and the importance of selecting models that support advanced tooling features.

In summary, the OntoLogX methodology delivers substantial gains in accuracy, completeness, and semantic faithfulness across a range of \glspl{llm}, demonstrating its effectiveness in guiding model outputs toward high-quality, ontology-compliant \gls{kg} construction. Small and inexpensive models are also viable, especially code-specific ones.

%% file: media/listings/baseline-prompt.tex
\begin{table}
\caption{Lines added to the prompt in \Cref{lst:main-prompt} to compose the baseline prompt for \gls{kg} generation.}
\label{lst:baseline-prompt}
\begin{minted}
[
frame=lines,
framesep=2mm,
baselinestretch=1,
fontsize=\footnotesize,
breaklines=true,
breaksymbolleft={},
breaksymbolright={},
]
{markdown}
# Output Format
The output graph must be in the following JSON format: {{json output format}}
Each node type has a specific set of allowed properties. The allowed properties for each node type are: {{properties schema}}
Each relationship type has a predefined source and target node type. The allowed relationships, formatted as (source type, relationship type, target type), are: {{triples}}
# Strict Compliance
Adhere to these rules strictly. Any deviation will result in termination.
\end{minted}
\end{table}

%% file: media/tables/llms.tex
\begin{table}
\caption{Comparison of \glspl{llm} used in experiments. The number of parameters indicates the size of the model.}
\label{tab:llms}
\renewcommand{\arraystretch}{1.2}
\begin{tabular}{@{}lllll@{}}
\toprule
Model & \# of parameters & License &  &  \\ \midrule
Claude 3.5 Haiku & Unknown, small size & Proprietary &  &  \\
Claude 3.5 Sonnet & Unknown, medium size & Proprietary &  &  \\
Llama 3.3 & 70 billions & Custom\tablefooturl{https://mistral.ai/static/licenses/MNPL-0.1.md}, permissive &  &  \\
Mistral Large 24.02 & 123 billions & Custom\tablefooturl{https://www.llama.com/license/}, permissive &  &  \\
Qwen 2.5 Coder & 14 billions & Apache 2.0 &  &  \\ \bottomrule
\end{tabular}
\end{table}

%% file: media/tables/model-performance.tex
\begin{table}
\caption{Model performance comparison between Baseline and OntoLogX. Given a row and a metric, bold values indicate the better mean performance.}
\centering
\resizebox{\textwidth}{!}{%
\renewcommand{\arraystretch}{1.2}
\begin{tabular}{@{}lllllllllllllllll@{}}
\toprule
\multicolumn{1}{c}{\multirow{3}{*}{Model}} & \multicolumn{4}{c}{Precision} & \multicolumn{4}{c}{Recall} & \multicolumn{4}{c}{F1 Score} & \multicolumn{4}{c}{G-Eval Score} \\ \cmidrule(l){2-17} 
\multicolumn{1}{c}{} & \multicolumn{2}{l}{Baseline} & \multicolumn{2}{l}{OntoLogX} & \multicolumn{2}{l}{Baseline} & \multicolumn{2}{l}{OntoLogX} & \multicolumn{2}{l}{Baseline} & \multicolumn{2}{l}{OntoLogX} & \multicolumn{2}{l}{Baseline} & \multicolumn{2}{l}{OntoLogX} \\ \cmidrule(l){2-17} 
\multicolumn{1}{c}{} & Mean & SD & Mean & SD & Mean & SD & Mean & SD & Mean & SD & Mean & SD & Mean & SD & Mean & SD \\ \midrule
Claude 3.5 Haiku & 0.778 & 0.006 & \textbf{0.857} & 0.007 & 0.453 & 0.004 & \textbf{0.820} & 0.007 & 0.573 & 0.004 & \textbf{0.838} & 0.006 & 0.714 & 0.008 & \textbf{0.750} & 0.007 \\
Claude 3.5 Sonnet & 0.753 & 0.006 & \textbf{0.859} & 0.005 & 0.514 & 0.003 & \textbf{0.858} & 0.007 & 0.611 & 0.002 & \textbf{0.859} & 0.005 & 0.728 & 0.012 & \textbf{0.767} & 0.007 \\
Llama 3.3 & 0.810 & 0.004 & \textbf{0.904} & 0.007 & 0.461 & 0.007 & \textbf{0.814} & 0.007 & 0.588 & 0.006 & \textbf{0.856} & 0.005 & 0.660 & 0.016 & \textbf{0.720} & 0.010 \\
Mistral Large & 0.803 & 0.011 & \textbf{0.811} & 0.019 & \textbf{0.406} & 0.012 & 0.286 & 0.015 & \textbf{0.539} & 0.011 & 0.423 & 0.015 & \textbf{0.637} & 0.025 & 0.418 & 0.030 \\
Qwen 2.5 Coder & 0.420 & 0.450 & \textbf{0.870} & 0.015 & 0.005 & 0.006 & \textbf{0.759} & 0.015 & 0.009 & 0.011 & \textbf{0.811} & 0.011 & 0.008 & 0.009 & \textbf{0.723} & 0.014 \\ \bottomrule
\end{tabular}%
}
\label{tab:model-performance}
\end{table}

%% file: parts/conclusions.tex
This paper introduces OntoLogX, a novel methodology for extracting semantically enriched \gls{cti} from cybersecurity logs using a combination of ontology-driven structured output and \glspl{llm}. The approach addresses the challenges of interpreting unstructured and ambiguous log entries by integrating ontological knowledge to guide the language model's reasoning and validate extracted data. The extracted information is then organized into an ontology-enriched graph database, facilitating future semantic analysis and querying.

The experimental evaluation, conducted using logs of various types extracted from a publicly available dataset, demonstrates the effectiveness of OntoLogX, with an intentional focus on extraction quality over processing speed. The results indicate that the OntoLogX-enhanced setup consistently outperforms a prompt-only baseline configuration across precision, recall, and G-Eval metrics for nearly all models tested. Among the tested models, \texttt{Claude 3.5 Sonnet} achieved the best overall performance. In addition, code-focused models such as \texttt{Qwen 2.5 Coder} showed promising results, highlighting the potential of task-specific architectures.

Looking forward, several directions will guide future work. A key area is the development of interactive interfaces that allow analysts to engage with the ontology-enriched knowledge graph directly. These interfaces will support querying, validating, and refining extracted intelligence, facilitating human-in-the-loop workflows that combine automation with expert control. In parallel, we aim to implement adaptive feedback mechanisms that allow the system to learn from analyst input and respond to evolving adversarial behaviors. This will improve the resilience and adaptability of the extraction pipeline over time.

Another development involves the extension and alignment of the OntoLogX ontology. This includes expanding and integrating logs from other domains such as endpoint detection, cloud infrastructure, and application layers, and formalizing mappings to widely adopted CTI standards like SEPSES, \gls{uco}, and \gls{misp}. Such efforts are critical to ensure interoperability with existing threat intelligence platforms and support seamless integration into operational pipelines.

Finally, we plan to investigate further the use of task-specialized \glspl{llm}, particularly those optimized for reasoning or code generation, such as \texttt{Qwen 3} and \texttt{DeepSeek Coder}. When properly guided, these models may offer superior performance in graph generation and correction.